\documentclass[preprint,showkeys,secnumarabic,amsfonts,showpacs,amsmath,amssymb]{revtex4}
\usepackage[dvips]{color}
\usepackage{array}
\usepackage{rotating}
\usepackage{epsfig}
\usepackage{amsmath}
\usepackage{graphicx}
\begin{document}
\title{Entropy and statefinder diagnosis in chameleon cosmology }

\author{H. Farajollahi}
\email{hosseinf@guilan.ac.ir} \affiliation{Department of Physics,
University of Guilan, Rasht, Iran}
\author{A. Salehi}
\email{a.salehi@guilan.ac.ir} \affiliation{Department of Physics,
University of Guilan, Rasht, Iran}
\author{F. Tayebi}
\email{ftayebi@guilan.ac.ir} \affiliation{Department of Physics,
University of Guilan, Rasht, Iran}
\date{\today}

\begin{abstract}
 \noindent \hspace{0.35cm}

In this paper, the generalized second law (GSL) of thermodynamics and entropy is revisited in the context of cosmological models with bouncing behavior such as
 chameleon cosmology where the boundary of the universe is assumed to be enclosed by the dynamical apparent horizon. From a thermodynamic point of view, to link between thermodynamic and geometric parameters in cosmological models, we introduce "entropy rate of change multiplied by the temperature" as a model independent thermodynamic state parameter together with the well known $\{r,s \}$ statefinder to differentiate the dark energy models.

\end{abstract}


\maketitle

\section{introduction}

Recent observations of high redshift type Ia
supernovae, the surveys of clusters of galaxies, Sloan digital sky survey (
SDSS) and Chandra X--ray observatory reveal the universe accelerating expansion
and that the density of matter is very much less than the critical density \cite{Reiss}. In addition, the
observations of Cosmic Microwave Background (CMB)
anisotropies indicate that the universe is flat and the total energy
density is very close to the critical one \cite{Bennett}.

The observations strongly indicates that the universe
presently is dominated by a smoothly distributed and slowly
varying dark energy (DE) component. A dynamical equation of state ( EoS) parameter that is connected directly to the evolution of the energy density in the universe and indirectly to the expansion of the Universe can be regarded as a suitable parameter to explain the acceleration and the origin of DE \cite{faraj1}--\cite{Setare}. In scalar-tensor theories \cite{Sahoo}--\cite{Nojiri3}, interaction of the scalar field with matter ( for example in chameleon cosmology) can be used to interpret the late time acceleration and smoothly varying EoS parameter \cite{Setare1}--\cite{Dimopoulos}.

Motivated by the black hole physics, it was realized that there is a profound connection
between dynamic and thermodynamic of the unvierse (see for example \cite{Rindler}--\cite{relation between dynamics and thermodynamics}). In particular, the validity of the GSL {\cite{GSL}} which state that entropy of the fluid
inside the horizon plus the entropy associated with the apparent horizon do not decrease
with time, has been the subject of many studies. In here, the GSL is discussed for models such as chameleon cosmology with bouncing behavior.

In order to differentiate between cosmological models, a sensitive
and robust geometric diagnostic for dark energy models is proposed by Sahni et al. \cite{sahni} that makes use of parameter pair $\{r, s\}$, the so-called "statefinder". It probes the expansion dynamics of the universe through higher derivatives of the expansion factor as a natural companion to the deceleration parameter. In this work we relate the geometric properties of the cosmological models with the thermodynamic one via statefinder parameters. The "entropy rate of change times the temperature" is defined in terms of the second derivative of the scale factor of the universe, instead of deceleration parameter. This parameter is related to the statefinder parameter, and together are adopted to differentiate between dark energy models.

 Section one is devoted to the model independent, thermodynamic formulation of the cosmological models in relation to the dynamical parameters. In section two, we derive the field equations for chameleon model and examine its thermodynamic behavior. The statefinder parameter from a thermodynamic point of view is presented in section three followed by its application in chameleon model and a summary.

\section{GSL and entropy}

In the following we make two assumptions: 1) in addition to the entropy of the universe inside the horizon, an entropy is associated to the apparent/event horizon. 2) with the local equilibrium hypothesis, since the temperature of the fluid is almost equal to the apparent horizon temperature, the
energy would not spontaneously flow between the horizon and the fluid inside the horizon.

According to the GSL in an expanding universe, entropy of the viscous dark energy (DE), dark matter (DM) and radiation
inside the horizon plus the one associated with the apparent horizon do not decrease
with time. In general, there are two approaches to validate the GSL on apparent/event horizons: 1) by using first law of thermodynamics and find entropy relation on the horizons i.e., \cite{horizon.first-law1},\cite{horizon.first-law.Bousso},
\begin{eqnarray}\label{first-law1}
T_{h}dS_{h}=-dE_{h}=4 \pi R_{h}^{3} H T_{\mu\nu}\kappa^{\mu}\kappa^{\nu}dt=4 \pi R_{h}^{3} H (\rho_{eff}+p_{eff})dt,
\end{eqnarray}
where $\kappa^{\mu} = (1, -Hr, 0, 0)$ are the (approximate) Killing vector ( the generators of the horizon), or the
future directed ingoing null vector field  and $"h"$ stands for the horizon. 2) in the field equations, by employing the horizon entropy and temperature formula on the horizon,
\begin{eqnarray}
S_{h}=\pi R_{h}^{2},\label{entropy}\\
T_{h}=\frac{1}{2\pi R_{h}}\label{temp.1}.
\end{eqnarray}
Note that only on the apparent horizon the two approaches are equivalent \cite{GSL in BD}. According to the recent observational data from type Ia Supernovae in an accelerating universe, the enveloping surface should be the apparent
horizon rather than the event one \cite{AH instead EH}. Therefore, from now on, we assume that the universe is enclosed by the dynamical apparent horizon with the radius given by $R_{h}=\frac{1}{\sqrt{H^{2}}}$ in a flat FRW universe \cite{R_AH}.

By the horizon entropy and temperature given in equations (\ref{entropy}) and (\ref{temp.1}),
the dynamics of the entropy on the apparent horizon is \cite{GSL},
\begin{eqnarray}
\dot{S_{h}}=2\pi R_{h} \dot{R_{h}}.
\end{eqnarray}
Also, from the Gibbs equation, the entropy of the universe inside the horizon can be related
to its effective energy density and pressure in the horizon with,
\begin{eqnarray}\label{Gibb's eq.1}
TdS_{in}=p_{eff}dV+d(E_{in}),
\end{eqnarray}
where $S_{in}$ is the internal entropy within the apparent horizon and $p_{eff}$ is the effective pressure in the model. If there is no energy exchange between outside and inside of the apparent horizon, thermal equilibrium realizes that $T = T_{h}$. Hence, the expression for internal energy can be written as $E_{in} = \rho_{eff}V$, with $V = \frac{4}{3}\pi R_{h}^{3}$. From equation (\ref{Gibb's eq.1}), by using Friedmann equation in the cosmological models and doing some algebraic manipulations we find that the rate of change of the internal entropy, horizon entropy and total entropy are respectively,
\begin{eqnarray}
\dot{S_{in}}&=&12\pi^2 R_h^2H(1+\omega_{eff})(1+3\omega_{eff}),\label{sdotin}\\
\dot{S_{h}}&=&24\pi^2 R_h^2H(1+\omega_{eff}),\label{sdoth}\\
\dot{S_{t}}&=&36\pi^2 R_h^2H(1+\omega_{eff})^2.\label{sdott}
\end{eqnarray}
For the rate of change of the internal entropy, equation (\ref{sdotin}), we find that for an expanding universe with acceleration, $H>0$, and $-1 <\omega_{eff}<-1/3$ in quintessence era, $\dot{S_{in}}< 0$. On the other hand, in phantom era, $\omega_{eff}<-1$, and decelerating universe, $\omega_{eff}>-1/3$, we obtain $\dot{S_{in}}>0$. From equation (\ref{sdoth}), it can be seen that again in an decelerating expanding universe and when $\omega_{eff}>0$, then $\dot{S_{h}}\geq 0$. Otherwise, $\dot{S_{h}}\leq 0$. Finally, in equation (\ref{sdott}), the sign of the total rate of change of the entropy, $\dot{S_{t}}$, for an expanding universe is independent of EoS parameter.

\section{Chameleon Model}

In this section, we consider the chameleon gravity in the presence of matter with the action given by,
\begin{eqnarray}\label{action}
S=\int[\frac{R}{16\pi
G}-\frac{1}{2}\phi_{,\mu}\phi^{,\mu}+V(\phi)+f(\phi)L_{m}]\sqrt{-g}dx^{4},
\end{eqnarray}
where $R$ is Ricci scalar, $G$ is the newtonian constant gravity
and $\phi$ is the chameleon scalar field with the potential
$V(\phi)$. Unlike the usual Einstein-Hilbert action, the matter
Lagrangian $L_{m}$ is modified as $f(\phi)L_{m}$, where $f(\phi)$ is
an analytic function of $\phi$. This last term in Lagrangian brings about the nonminimal
interaction between matter and chameleon field. The variation of the action (\ref{action}) with respect to the metric tensor components in a spatially flat FRW  cosmology yields the field equations,
\begin{eqnarray}
&&3H^{2}=\rho_{m}+\frac{1}{2}\dot{\phi}^{2}+V(\phi),\label{fried1}\\
&&2\dot{H}+3H^2=-\gamma\rho_{m}-\frac{1}{2}\dot{\phi}^{2}+V(\phi),\label{fried2}
\end{eqnarray}
where we put  $8\pi G=c=\hbar=1$ and $\rho_{m}=\rho_{m}(a,\phi)$. The dots means derivatives with respect to the cosmic time $t$. The energy density $\rho_{m}$ is the matter energy density in the universe. Also
variation of the action (\ref{action}) with respect to the scalar field  $\phi$ provides the wave
equation for the chameleon field as,
\begin{eqnarray}\label{phiequation}
\dot{\phi}(\ddot{\phi}+3H\dot{\phi})=-\dot{V}-(ln X)'\rho_{m},
\end{eqnarray}
where we have introduced $X(\phi)=f(\phi)^{1-3\gamma}$. The parameter $\gamma$ is the barotropic index
of the background fluid (DM). From equations (\ref{fried1}), (\ref{fried2}) and (\ref{phiequation}), one can easily arrive at the relation,
\begin{eqnarray}\label{conserv}
\dot{\rho_{m}}+3H(1+\gamma)\rho_{m}=(ln X)'\rho_{m}\dot{\phi},
\end{eqnarray}
which for $\rho_{m}(a,\phi)=\rho_{0m}(a)f(\phi)$ readily integrates to yield
\begin{eqnarray}
\rho_{m}(a,\phi)=MX^{-1}a^{-3(1+\gamma)}=Mf^{-3(1+\gamma)}a^{-3(1+\gamma)},
\end{eqnarray}
with $M$ as a constant of integration. From equations (\ref{fried1}) and (\ref{fried2}) and in comparison with the standard friedmann equations we identify $\rho_{eff}$ and $p_{eff}$ as,
\begin{eqnarray}\label{roef}
\rho_{eff}\equiv\rho_{m}+\frac{1}{2}\dot{\phi}^{2}+V(\phi).
\end{eqnarray}
\begin{eqnarray}\label{pef}
p_{eff}\equiv\gamma\rho_{m}+\frac{1}{2}\dot{\phi}^{2}-V(\phi),
\end{eqnarray}
with the equation of state, $p_{eff}=\omega_{eff}\rho_{eff}$.

For the cosmological models including chameleon cosmology with a possibility of bouncing behavior, the total entropy rate of change can be negative, thus, GSL is violated when the universe is in contraction era. To study the thermodynamic behavior of the chameloen model at late time, the structure of the dynamical system is revisited by taking into account the following dimensionless variables,
\begin{eqnarray}\label{defin}
 \Omega_{\phi_{K.E}}=\frac{\dot{\phi}^2}{6 H^2},\ \ \Omega_{\phi_{P.E.}}=\frac{V}{3H^{2}} ,\ \ \Omega_{m}f={\frac{\rho_{m}f}{3 H^{2}}}.
\end{eqnarray}
In order to close the system of equations we make the following ansatz: We consider that $f(\phi)=f_{0}\exp{(\delta_{1}\phi)}$ and   $V(\phi)=V_{0}\exp{(\delta_{2}\phi)}$ where $\delta_{1}$ and $ \delta_{2} $ are  dimensionless constants characterizing the slope of potential $V(\phi)$ and $f(\phi)$. There are no priori physical motivation for these choices, so it is only purely phenomenological which leads to the desired behavior of phantom crossing. Using equations (\ref{fried1})-(\ref{conserv}), the evolution equations of these variables become,
\begin{eqnarray}
\frac{d\Omega_{\phi_{K.E}}}{dN}&=&2\Omega_{\phi_{K.E}}[-3+\frac{3}{2}(1+\gamma)\Omega_{m}f
+3\Omega_{\phi_{K.E}}]-\sqrt{6\Omega_{\phi_{K.E}}}[\delta_{1}(1-3\gamma)\Omega_{m}f+
\delta_{2}\Omega_{\phi_{P.E}}],\label{x1}\\
\frac{d\Omega_{\phi_{P.E}}}{dN}&=&\Omega_{\phi_{P.E}}[\sqrt{6}\delta_{2}\sqrt{\Omega_{\phi_{K.E}}}+
(1+\gamma)\Omega_{m}f
+6(\Omega_{m}f)^{2}],\label{z1}\\
\frac{d(\Omega_{m}f)}{dN}&=&\Omega_{m}f[-3(1+\gamma)+3(1+\gamma)\Omega_{m}f+
\sqrt{6}\delta_{1}(1-3\gamma)\sqrt{\Omega_{\phi_{K.E}}}
+6\Omega_{\phi_{K.E}}],\label{y1}
\end{eqnarray}
where $N = ln (a)$. The Friedmann constraint equation (\ref{fried1}) can be written as
\begin{eqnarray}\label{constraint}
\Omega_{D}+\Omega_{m}f=1 ,
\end{eqnarray}
where $\Omega_{D}=\Omega_{\phi_{K.E}}+\Omega_{\phi_{P.E}}$.
In term of the new dynamical variables we also have,
\begin{eqnarray}
\frac{\dot{H}}{H^{2}}=\frac{-1}{2}[3(1+\gamma)\Omega_{m}f+6\Omega_{\phi_{K.E}}].
\end{eqnarray}
 which can be used later in deriving other dynamical parameters. Using the Friedmann constraint (\ref{constraint}), equations (\ref{x1})-(\ref{y1}) reduces to
\begin{eqnarray}\label{x2}
\frac{d\Omega_{\phi_{K.E}}}{dN}&=&2\Omega_{\phi_{K.E}}[3(1-\gamma)+\sqrt{6}(1-3\gamma)
\delta_{1}\Omega_{m}f
-6(1-\Omega_{m}f-\Omega_{\phi_{K.E}})\nonumber\\
&-&3(1-\gamma)\sqrt{\Omega_{\phi_{K.E}}}],\\
\frac{d(\Omega_{m}f)}{dN}&=&\Omega_{m}f[-\frac{3(1-\gamma)}{2}\sqrt{\Omega_{\phi_{K.E}}}-
3(1-\Omega_{m}f-\Omega_{\phi_{K.E}})
+\frac{\sqrt{6}}{2}\delta_{2}]\nonumber\\
&-&\frac{\sqrt{6}(1-3\gamma)\delta_{1}}{2}\sqrt{\Omega_{\phi_{K.E}}}-\frac{\sqrt{6}\delta_{2}}{2}
(1-\Omega_{\phi_{K.E}}).\label{y2}
\end{eqnarray}
It is more convenient to work with the equations (\ref{x2}) and (\ref{y2}) than equations (\ref{x1})-(\ref{y1}). With respect to these equations, the total entropy rate of change becomes,
\begin{eqnarray}
\dot{S_{t}}=8\pi(\frac{\dot{H}}{H^2})^2=2\pi[3(1+\gamma)\Omega_{m}f+6\Omega_{\phi_{k}}]^2.
\end{eqnarray}
With numerical computation, in Fig.1, the dynamics of total entropy rate of change and EoS parameter with respect to the redshift $z$ for $\gamma=0, 1/3$ are plotted.\\

\begin{tabular*}{2.5 cm}{cc}
\includegraphics[scale=.35]{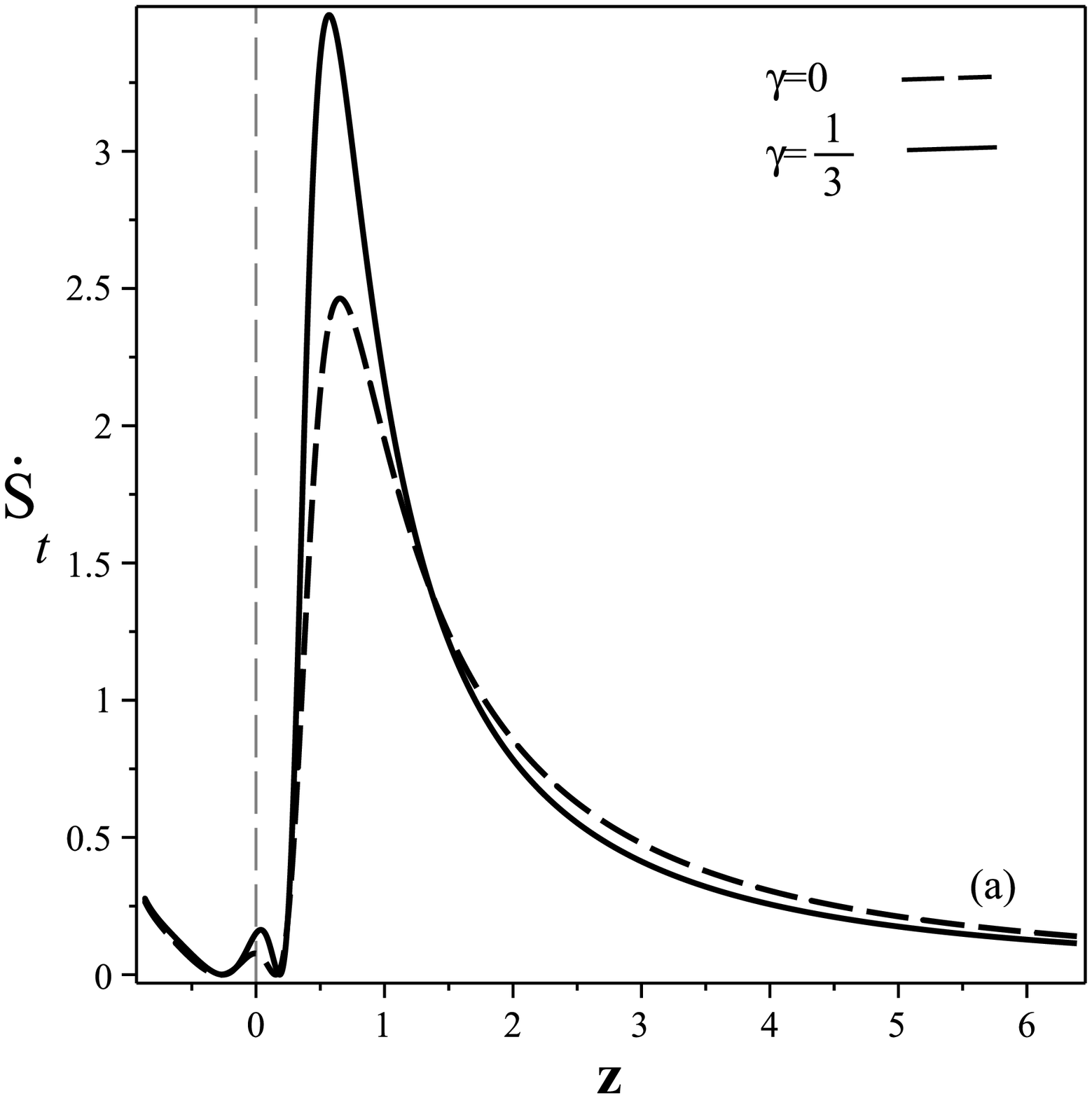}\hspace{0.1 cm}\includegraphics[scale=.4]{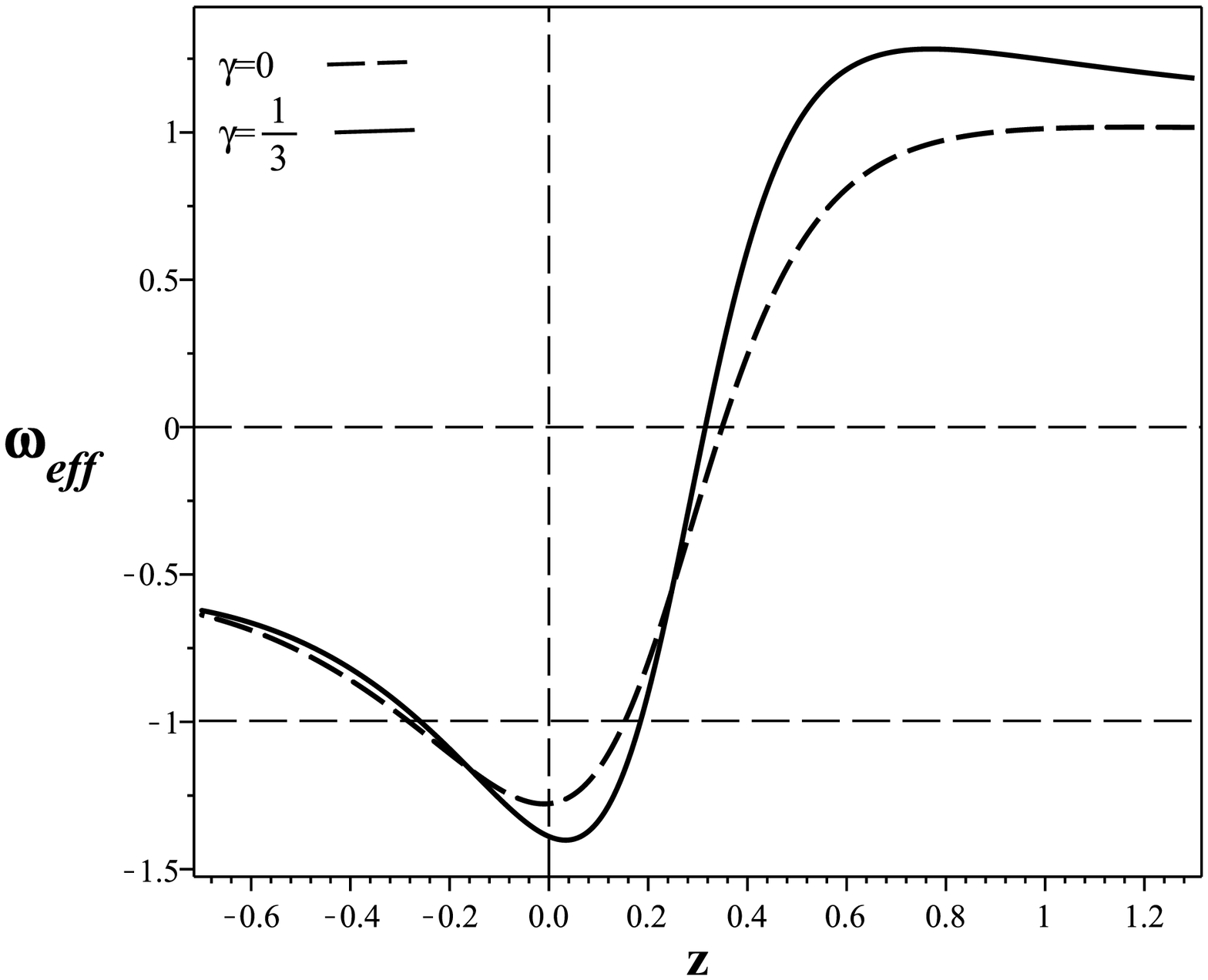}\hspace{0.1 cm}\\
Fig. 1:  The trajectories of the total entropy rate of change evolution and \\ EoS parameter against redshift  $z $ for $\gamma=0, 1/3$.\\
ICs: $\Omega_{\phi_{K.E}}|_0=-\frac{36}{6}$, $\Omega_{m}f|_0=-\frac{1}{3}$.
\end{tabular*}\\

From the graphs, at higher redshifts, the EoS parameter is positive and the universe decelerates. The $\dot{S_{t}}$ gradually increases to its local maximum value when $\omega_{eff}$ is maximum at $z\simeq 0.7$ in a decelerating expansion universe. It then sharply decreases to its local minimum value, ($\dot{S_{t}}=0$), correspond to the phantom crossing in the past with a fast transition from a decelerating expansion to an accelerating one. By entering the phantom era, ($\omega_{eff}<-1$), $\dot{S_{t}}$ starts to increase to its local maximum value when EoS parameter reaches its minimum. While EoS parameter continue its journey from phantom era to quintessence era passing once again the phantom barrier in future, the $\dot{S_{t}}$ function reaches its local minimum and then increases. Since $\dot{S_{t}}$ function vanishes at its local minimums where $\ddot{S_{t}}=0$ at the phantom crossing, the entropy function has points of inflexion at crossing points.

\section{Statefinder diagnosis}

In general, for the statefinder of the cosmological models, we define $u=\dot{S_{t}}T=8\pi(\frac{\dot{H}}{H^2})^2$, as a thermodynamic parameter, instead of the geometric deceleration parameter. Thus, the statefinder parameter $\{r,s\}$ as the geo-thermodynamics parameter in terms of $u$ become,
\begin{eqnarray}
r=\frac{\ddot{H}}{H^{3}}-\frac{3}{2}(\frac{u}{2\pi})^{1/2}+1, \mbox{}\ \ s=\frac{2(r-1)}{3(\frac{u}{2\pi})^{1/2}-3}.
\end{eqnarray}
We therefore have in the three cosmological scenarios, $\{u, r, s \} = \{18\pi, 1, 0 \}$ as a fixed point for $SCDM$ state, $\{u, r, s \} = \{2\pi, 1, 0 \}$ as a fixed point for $\small{LCDM}$ state, and $\{u, r, s \} = \{18\pi, 1, 1\}$ as a fixed point for $SS$ state.
In terms of the dimensionless variables in the chameleon model we obtain,
\begin{eqnarray}
r&=&3\sqrt{\Omega_{\phi_{K.E}}}(3+\frac{\sqrt{6}\Omega_{m}f}{8}\delta_{1}(1-\gamma)(1-3\gamma)
+\sqrt{6}\delta_{2}\Omega_{\phi_{P.E}})+\frac{9\Omega_{m}f\gamma}{2}(1+\gamma)+1,\\
s&=&\frac{-6\Omega_{\phi_{K.E}}-(1+\gamma)\Omega_{m}f(3\gamma+\delta_{1}\frac{\sqrt{6}}{4}(1-3\gamma)
\sqrt{\Omega_{\phi_{K.E}}})
-2\sqrt{6}\delta_{2}\sqrt{\Omega_{\phi_{K.E}}}\Omega_{\phi_{P.E}}}{6[1+(1+\gamma)\Omega_{m}f+
2\Omega_{\phi_{K.E}}]}.
\end{eqnarray}
We also have $u$ in terms of these variables as,
\begin{eqnarray}
u=2\pi[3(1+\gamma)\Omega_{m}f+6\Omega_{\phi_{K.E}}]^2.
\end{eqnarray}
Fig.3 shows the statefinder diagrams $\{u, z\}$, $\{u, r\}$ and $\{u, s\}$ evolutionary
trajectories for the chameleon model from a thermodynamic point of view. From the graphs, all the trajectories for $\gamma=0,1/3$ commence evolving from a critical point in the past and end their evolution to a second critical point in the future. The current values of the statefinder on the trajectories show that the universe currently is between LCDM state in the past and SS state in the future. There is thermodynamic stability during the universe expansion so that the universe cools down through the expansion. The universe transits from a state in the past to the final state in the future while not passing through the above states.

\begin{tabular*}{2.5 cm}{cc}
\includegraphics[scale=.3]{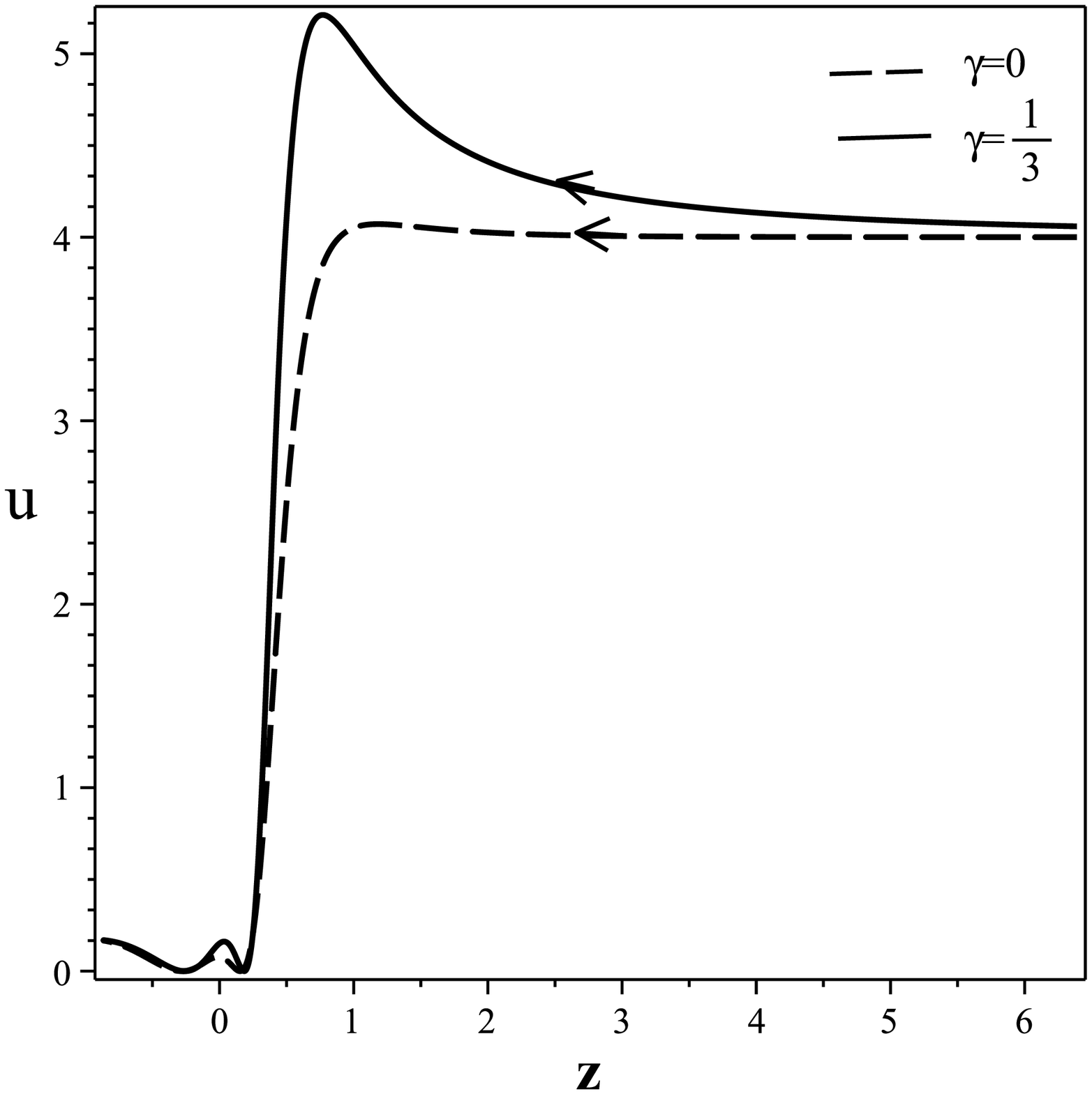}\hspace{0.1 cm}\includegraphics[scale=.3]{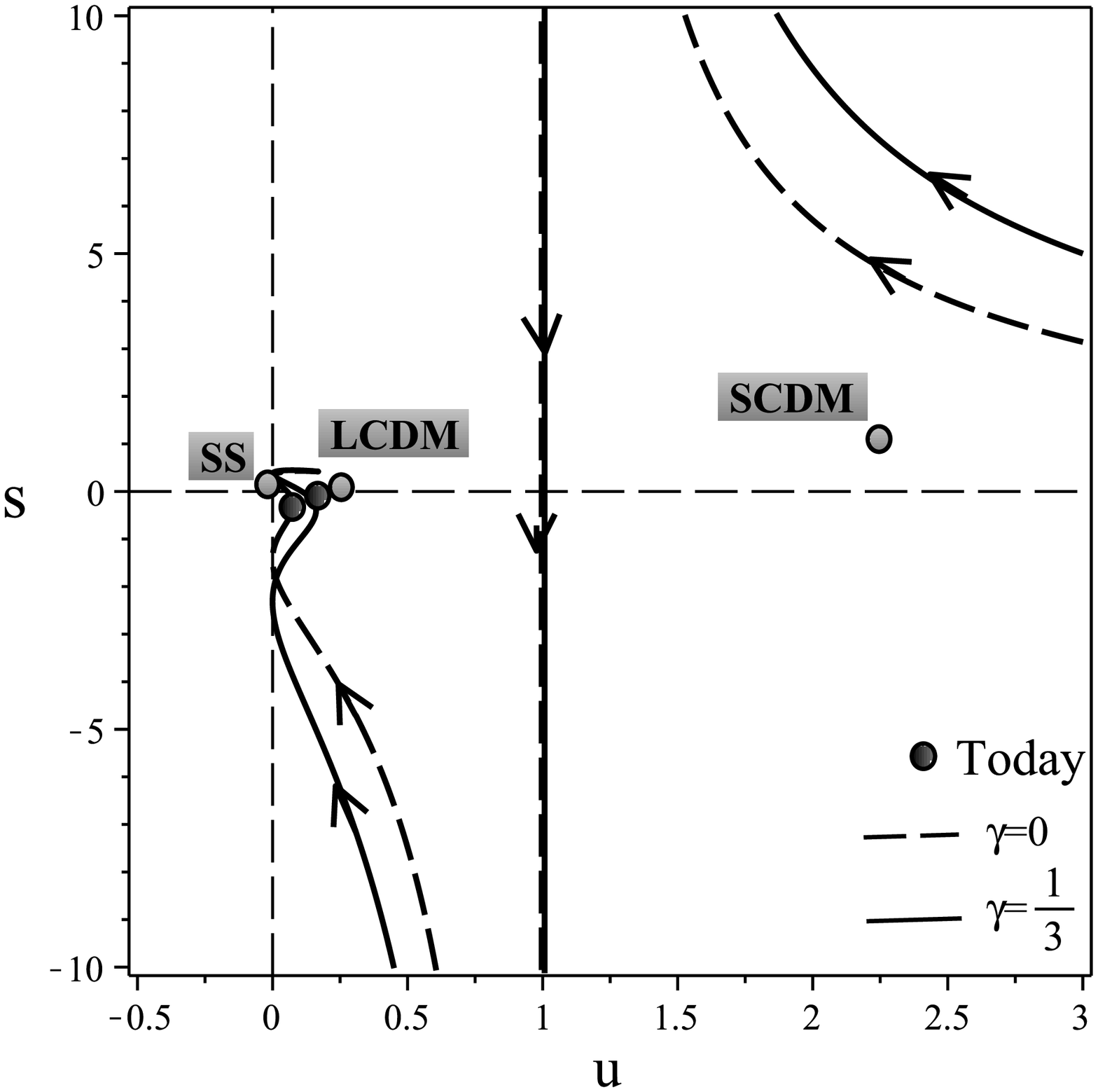}\hspace{0.1 cm}\includegraphics[scale=.3]{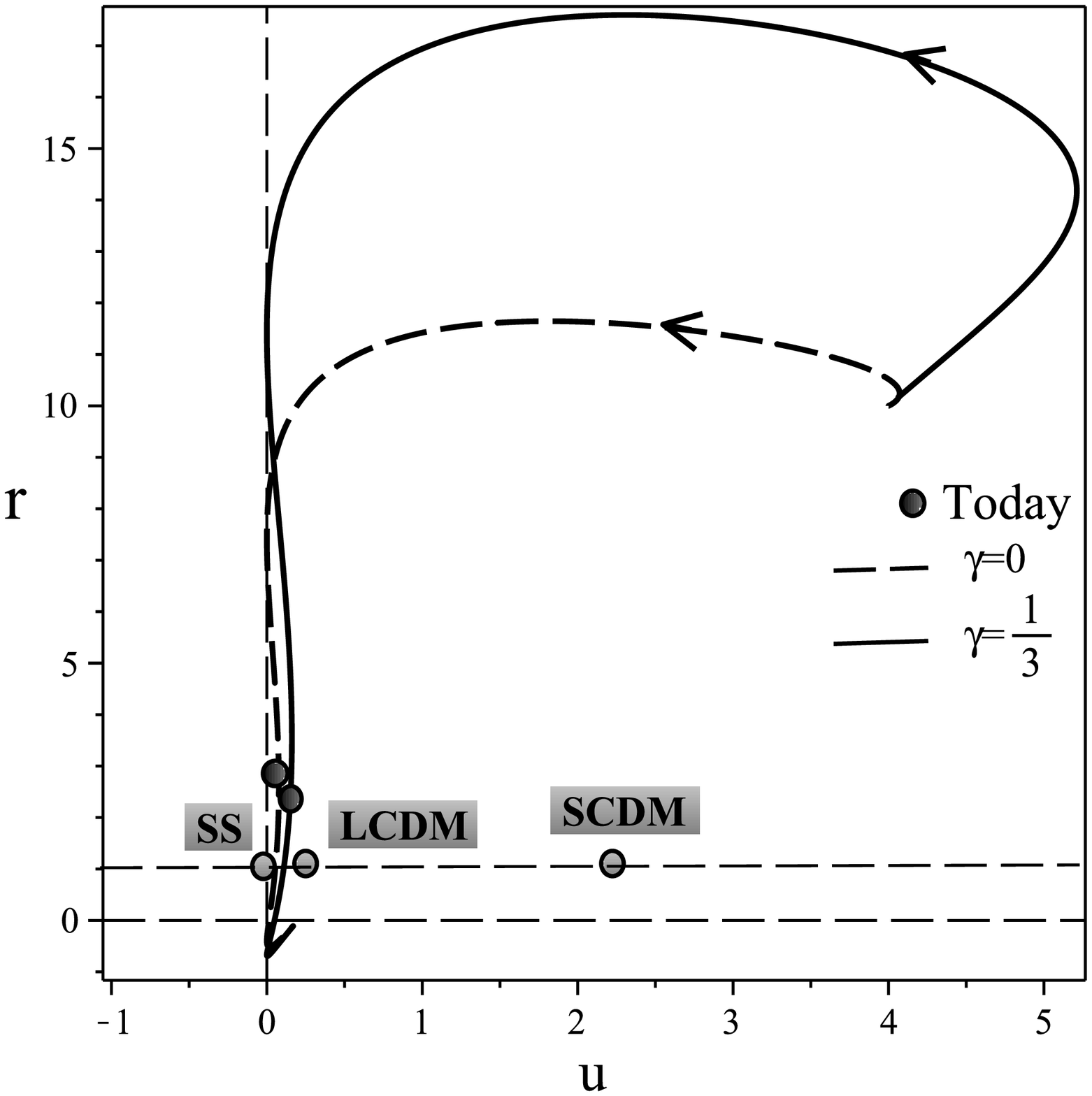}\hspace{0.1 cm}\\
Fig. 3:  The trajectories of the statefinder $\{r, u\}$ and $\{s, u\}$ and $u$ against redshift $z$,\\
for both $\gamma=0,1/3$. ICs: $\Omega_{\phi_{K.E}}|_0=-\frac{36}{6}$, $\Omega_{m}f|_0=-\frac{1}{3}$.
\end{tabular*}\\

In summary, we  realize that the total entropy of the universe in chameleon cosmology with bouncing behavior and effective EoS parameter increases/decreases with time in the expansion/contraction period thus GSl is violated in the contraction period. The total entropy rate of change as a model independent parameter is discussed for the model. For the chameloen model, the universe starts from a state in the past with $\dot{S}_t>0$, excluding the early early time when a possible bouncing occurred, and following its journey to another state in future again with positive $\dot{S}_t$. Phantom crossings in the past and future correspond to the $\dot{S}_{t}|_{min}=0$ when the entropy function has points of inflexion. We then introduce $\dot{S}_tT$ to link between geometric and thermodynamic behavior of the cosmological models, instead of deceleration parameter. This parameter together with the corresponding statefinder $\{r,s\}$ can be used to differentiate the dark energy models from a thermodynamic point of view.

\end{document}